\documentstyle[11pt,twoside,psfig,jltp]{article}

\title{Low-dimensional BEC}

\author{{{F.J. Sevilla$^{1}$,}} M. Grether$^{2}$, M. Fortes$^{1}$, M.
de Llano$^{3}$ O. Rojo$^{4}$, \\
M.A. Sol\'{\i}s$^{1}$ and A.A. Valladares$^{3}$}

\address{$^{1}$Instituto de F\'{\i}sica, UNAM, Apdo. Postal 20-364, 01000 M\'{e}xico, DF, Mexico \\ $^{2}
$Facultad de Ciencias, UNAM, 04510 M\'{e}xico, DF, Mexico \\ $^{3} $%
Instituto de Investigaciones en Materiales, UNAM, 04510 M\'{e}xico, DF,
Mexico \\ $^{4}$PESTIC, Secretar\'{\i}a Acad\'{e}mica \& CINVESTAV, IPN,
04430 M\'{e}xico, DF, Mexico}

\runninghead{F.J. Sevilla {\it et al.}}{Low-dimensional BEC}

\begin{document}
 
\vspace{-1.0cm}

\begin{abstract}
The Bose-Einstein condensation (BEC) temperature $T_{c}$ of Cooper pairs
(CPs) created from a general interfermion interaction is determined for a 
{\it linear}, as well as the usually assumed quadratic, energy {\it vs }%
center-of-mass momentum dispersion relation. {\it \ }This\ explicit $T_{c}$
is then compared with a widely applied {\it implicit} one of Wen \& Kan
(1988) in $d=2+\epsilon $ dimensions, for small $\epsilon $, for a geometry
of an infinite stack of parallel (e.g., copper-oxygen) planes as in, say, a
cuprate superconductor, and with a new result for linear-dispersion CPs. The
implicit formula gives $T_{c}$ values only slightly lower than those of the
explicit formula for typical cuprate parameters. 

\end{abstract}
\maketitle


Bose-Einstein condensation (BEC) of Cooper pairs (CPs) can lead to a phase
transition (even in 2D) in any many-fermion system dynamically capable of
forming CPs. This transition could be the origin of ``exotic''
superconductivity in the quasi-2D cuprates and in the quasi-1D
organo-metallic (Bechgaard) salts, as well as of the superfluidity in liquid 
$^{3}$He or in trapped Fermi gases in 3D.

The familiar BEC formula for the transition temperature is 
\begin{equation}
T_{c}\simeq 3.31\hbar ^{2}n_{B}^{2/3}/m_{B}k_{B},  \label{BEC}
\end{equation}
with $n_{B}$ the number density of bosons of mass $m_{B}$ and $k_{B}$ the
Boltzmann constant. This is a special case of the more general expression 
\cite{pla} valid for any space dimensionality $d>0$ and any boson dispersion
relation $\varepsilon _{K}=C_{s}\,K^{s}$ with $s>0$ and $C_{s}$ constant,
given by the {\it explicit} $T_{c}$-formula 
\begin{equation}
T_{c}=\frac{C_{s}}{k_{B}}\left[ \frac{s\,\Gamma (d/2)\,(2\pi )^{d}n_{B}}{%
2\pi ^{d/2}\,\Gamma (d/s)g_{d/s}(1)}\right] ^{s/d}.  \label{gentc}
\end{equation}
If $\mu (T)$ is the boson chemical potential and $e^{\mu (T)/k_{B}T}\equiv z$
the fugacity, $g_{\sigma }(z)\equiv \sum_{l=1}^{\infty }z^{l}/l^{\sigma }$
are the Bose integrals. For $z=1$ and $\sigma \geq 1$ the $g_{\sigma }(1)$
is just $\zeta (\sigma )$, the Riemann Zeta-function of order $\sigma $
which is finite for $\sigma >1$ and infinite for $\sigma =1$, while the
series $g_{\sigma }(1)$ diverges for all $\sigma \leq 1$. \ For $s=2$, $%
C_{2}=\hbar ^{2}/2m_{B}$, and since $\zeta (3/2)\simeq 2.612$, this leads to
the usual BEC $T_{c}$-formula (\ref{BEC}). \ Since $g_{d/2}(1)$ diverges for
all $d/2\leq 1$, $T_{c}=0$ for all $d\leq 2$. This follows from the boson
number equation 
\begin{equation}
N=N_{0}(T)+\sum_{{\bf K}\neq 0}\left[ e^{\{\varepsilon _{K}-\mu
(T)\}/k_{B}T}-1\right] ^{-1}  \label{bec}
\end{equation}
where $N_{0}(T)$ is the number of bosons in the $K=0$ state. At $T=T_{c}$
both $N_{0}(T_{c})$ {\it and }the boson chemical potential $\mu (T_{c})$
virtually vanish so that replacing 
\begin{equation}
\sum_{{\bf k\neq 0}}\quad {\longrightarrow }\quad (L/2\pi
)^{d}\int_{0^{+}}d^{d}k=(L/2\pi )^{d}{\frac{2\pi ^{d/2}}{\Gamma (d/2)}}%
\int_{0^{+}}^{\infty }dk\,k^{d-1}  \label{replace}
\end{equation}
in (\ref{bec}) eventually yields (\ref{gentc}) where $n_{B}\equiv N/L^{d}$.

The fact that a CP can have a {\it linear }($s=1$), as opposed to the usual
quadratic ($s=2$), dispersion relation was mentioned as far back as 1964 by
Schrieffer \cite{sch64}, p. 33, for the BCS model interaction in 3D. This
was recently confirmed \cite{condmat}\ to be the case for both 2D and 3D\
under a very general interfermion interaction for any coupling provided the
fermion number-density is nonzero, i.e., in the presence of a Fermi sea. The
CP dispersion relation becomes quadratic{\it \ only in the extremely dilute }%
(or vacuum){\it \ limit} where the CPs are just the so-called ``local
pairs.'' For any sizeable fermion density the nonnegative CP {\it excitation
energy} $\varepsilon _{K}\equiv \Delta _{0}-\Delta _{K}$ behaves like $%
\simeq a(d)\hbar v_{F}K$, where $\Delta _{K}$ (not to be confused with the
BCS gap $\Delta $) is the (positive) binding energy of a CP of
center-of-mass momentum (CMM) $\hbar K$, $v_{F}\equiv \hbar k_{F}/m$ and $%
k_{F}$ the Fermi velocity and wavenumber, respectively, $m$ the fermion
effective mass, while\ $a(d)\equiv $ $2/\pi $ and $1/2$ in 2D and 3D,
respectively, precisely as established \cite{physc} previously for the BCS
model interaction. For linear dispersion $s=1$, $C_{1}=$ $a(d)\hbar v_{F}$, $%
T_{c}=0$ from (\ref{gentc}) for all $d\leq 1$ only---and $T_{c}>0$ for all $%
\ d>1$, which is precisely the range of dimensionalities for all known
superconductors if one includes the quasi-1D organo-metallic Bechgaard salts 
\cite{quai1d}. Using the interpolation $a(d)=(7/2-6/\pi )+(8/\pi
-13/4)d+(3/4-2/\pi )d^{2}$, which correctly reduces to $1,$ $2/\pi $ and $1/2
$ in 1D, 2D and 3D, respectively, Fig. 1 graphs (\ref{gentc}) for $s=1$ and $%
2$ (in units of the Fermi temperature $T_{F}$) {\it vs} $d$---if one
imagines all the fermions in the initial, interactionless many-fermion
system paired into CPs of mass $m_{B}=2m$. The particle number density of
the original fermions is $n\equiv k_{F}^{d}/2^{d-2}\pi ^{d/2}d\;\Gamma (d/2)$
and equals $2n_{B}$, and we have used $\hbar ^{2}k_{F}^{2}/2m=\frac{1}{2}%
mv_{F}^{2}=E_{F}\equiv k_{B}T_{F}$. These curves are {\it upper bounds} to
the $T_{c}$\ from a more realistic model \cite{Pret} where chemical
equilibrium allows only a {\it fraction} of all fermions to be actually
bound into pairs.
\noindent
\begin{figure}[tbh]
\begin{minipage}[b]{2.85in}
%\begin{figure}[tbh]
\centerline{\psfig{file=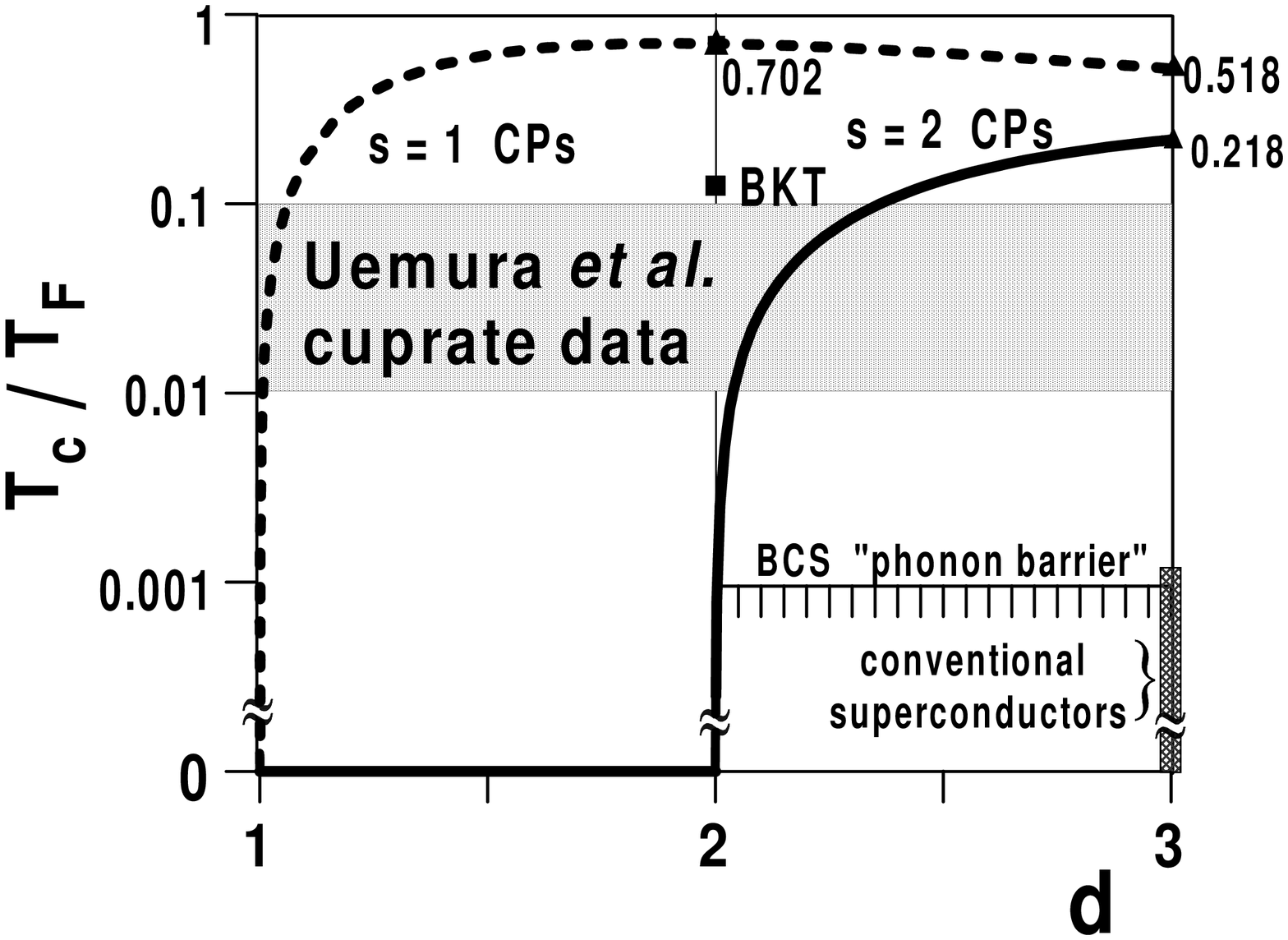,height=3.7in,width=2.9in}}
%
%\centerline{\framebox[3in]{\rule[1.125in]{0in}{1.125in}}}
%\makebox[5in]{\rule[1.125in]{0in}{1.125in}}
%\label{fig:tau2} 
%\end{figure} 
\end{minipage}\hfill
\begin{minipage}[b]{1.3in} 
%\begin{figure}[tbh]
\centerline{\psfig{file=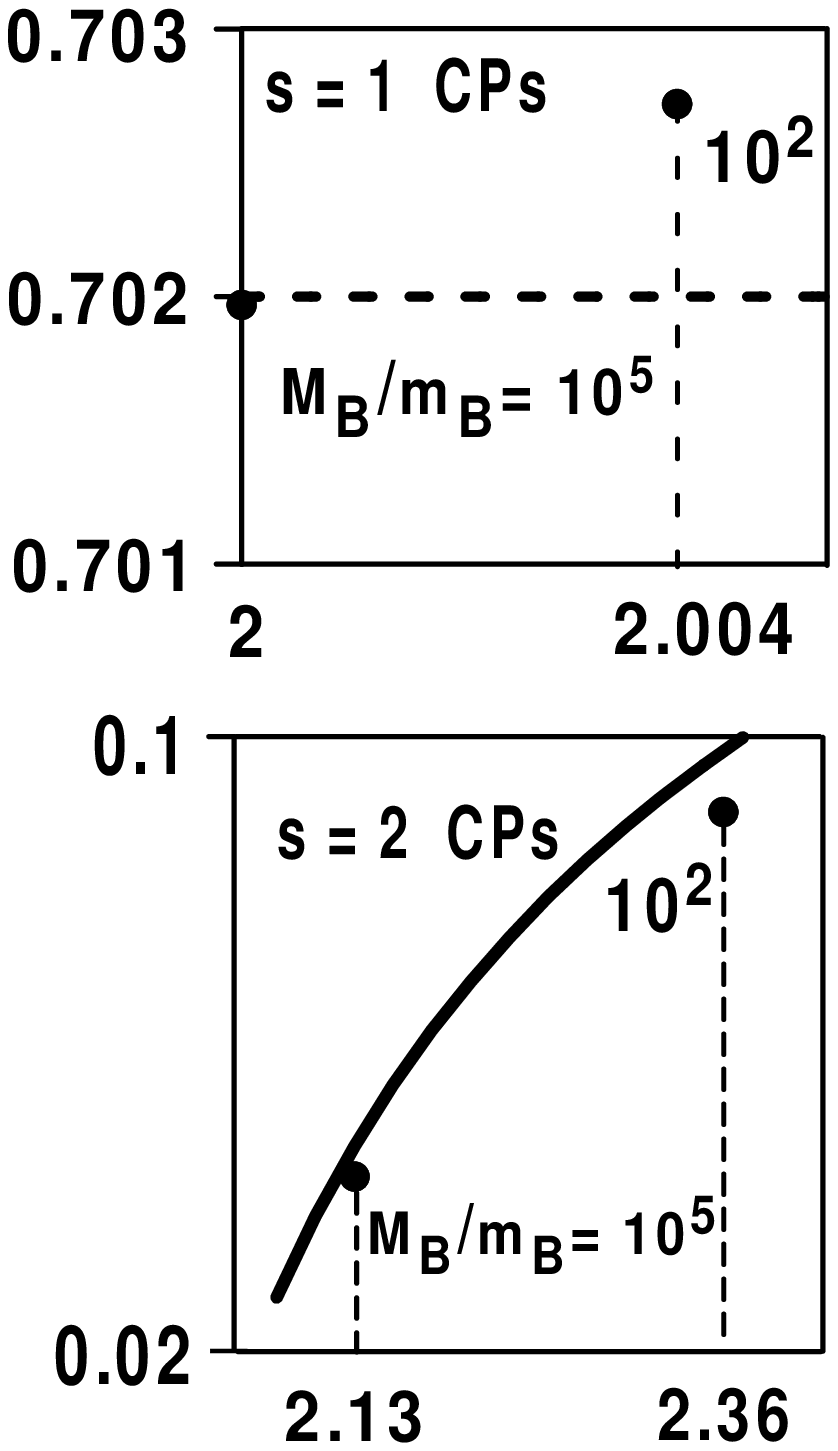,height=3.7in,width=2.9in}}
%
%\centerline{\framebox[3in]{\rule[1.125in]{0in}{1.125in}}}
%\makebox[5in]{\rule[1.125in]{0in}{1.125in}}
%\label{fig:tau2} 
%\end{figure}
\end{minipage}
\vspace{-3.5cm} 
\caption{ 
{\bf Left:} dimensionality, $d$, dependence of the critical BEC transition
temperature $T_{c}$ according to {\it explicit }(\ref{gentc}) (in units of
the Fermi temperature $T_{F}$) as explained in text. Lower (full) curve is
for $s=2$; upper (dashed) curve for $s=1$. \ Shaded areas refer to empirical
data from Ref. [7]. \ {\bf Right: }dots refer to results from {\it implicit }%
(\ref{WKeqn}) below, as explained just after (\ref{eps}). } 
\label{fig:tau2} 
\end{figure}

A rather general interfermion interaction is the $S$-wave attractive
separable potential whose double Fourier transform is 
\begin{equation}
V_{pq}=-(v_{0}/L^{2})g_{p}g_{q}.  \label{1}
\end{equation}
Here $L$ is the size of the ``box'' confining the many-fermion system, $%
v_{0}\geq 0$ is the interaction strength and $g_{_{p}}$ given, e.g., \cite
{18a} by $(1+p^{2}/p_{0}^{2})^{-1/2}$ where $p_{0}$ is the inverse range of
the potential. Hence, $p_{0}\rightarrow \infty $ implies $g_{_{p}}=1$ and
corresponds to the attractive contact (or delta) potential $%
V(r)=-v_{0}\delta ({\bf r})$, while $p_{0}=k_{F}$ implies a range of order
of the average interfermion spacing, etc. \ If $g_{p}=\theta (\hbar \omega
_{D}+\mu _{F}-p^{2}/2m)$, with $\theta (x)$ the unit step function, (\ref{1}%
) becomes the BCS model interaction where $\omega _{D}$ is the Debye
frequency and $\mu _{F}$ the fermionic chemical potential that becomes $E_{F}
$ for $T=0=v_{0}$.

Using a {\it renormalized CP equation} \cite{condmat} whose coupling depends 
{\it only }on the two-body binding energy $B_{2}$ \cite{miyake}, the CP
excitation energy $\varepsilon _{K}\equiv \Delta _{0}-\Delta _{K}$ for zero
range was obtained numerically \cite{condmat} as an exact curve that for
very small $B_{2}/E_{F}$ is virtually linear, i.e., $\varepsilon
_{K}\rightarrow 2\hbar v_{F}K/\pi $. \ It is {\it only} in the dilute limit (%
$v_{F}$ or $E_{F}\rightarrow 0$) that $\varepsilon _{K}$ tends
asymptotically to the exact quadratic $\hbar ^{2}K^{2}/2(2m)$ for any
coupling. \ Assuming $n_{B}=n/2$ and $m_{B}=2m$ to introduce the temperature
scale $T_{F}$ as before, Fig. 2 shows the BEC $T_{c}$'s of a pure gas of 
{\it unbreakable} CPs. \ Significantly, $T_{c}$ is {\it no longer} zero in
2D---as would be predicted in a BEC picture by a quadratic relation
appropriate for ``local-pair'' CPs in vacuum, a result that wrongly suggests
that BEC cannot apply for quasi-2D cuprate superconductors.

More accurate BEC $T_{c}$'s should include refinements such as non-$S$-wave
interactions, allowing for {\it unpaired fermions }in a more realistic
binary boson-fermion mixture model \cite{Pret}---and most importantly,
CP-fermion interactions that link \cite{Ren,Tolma} the BEC condensate
fraction temperature-dependence with that of the BCS fermionic energy gap,
among other corrections.

The linear dispersion relation of a CP {\it should not }be confused with the
linear dispersion of Anderson-Bogoliubov-Higgs (ABH) many-body excitation
phonon-like modes. Collective modes in a superconductor were studied since
the late 1950's by several workers. \ A more recent treatment for 1D, 2D and
3D is available \cite{Bel} which confirms the linear ABH form $\hbar v_{F}K/%
\sqrt{d}$ for $d$ =1, 2 or 3 in the zero-coupling limit. \ Our CPs are taken
as ``bosonic'' even though they do {\it not} obey (Ref. [2] p. 38) Bose
commutation relations. This is because for a given $K$ they have {\it %
indefinite} occupation number since for fixed $K$ there are (after the
thermodynamic limit) an indefinitely large number of allowed (relative
wavenumber) $k$ values, so that---for any{\sf \ }coupling and thus any
degree of overlap between them---CPs do in fact obey the Bose-Einstein
distribution from which BEC is determined. \ By contrast, ABH phonons (like
photons or plasmons, etc.) {\it cannot }suffer a BEC as their number is
always indefinite. The number of CPs, on the other hand, is {\it fixed} at
half the number of (pairable) fermions if all of these are imagined paired
at a given temperature and coupling.
\begin{figure}[tbh]
\centerline{\psfig{file=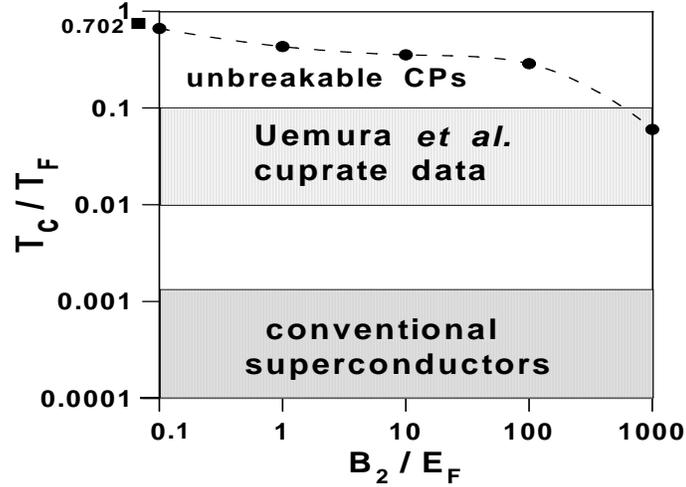,height=3.6in,width=3.5in}}
%
%\centerline{\framebox[3in]{\rule[1.125in]{0in}{1.125in}}}
%\makebox[5in]{\rule[1.125in]{0in}{1.125in}}
\vspace{-2.8cm}
\caption{
2D critical BEC temperatures $T_{c}$ (in units of $T_{F}$) for five coupling
values $B_{2}/E_{F}$ \ for a pure boson gas of unbreakable CPs, determined
by the $N_{0}(T_{c})=0=\mu (T_{c})$ solution of (\ref{bec}) inserting exact
numerical CP dispersion curves, using (\ref{replace}). The value 0.702
marked with a square corresponds to the $T_{c}/T_{F}$ value at zero coupling
(see also Ref. [1]) and contrasts with the well-known result $%
T_{c}/T_{F}\equiv 0$ in 2D for infinite coupling where $\varepsilon
_{K}=\hbar ^{2}K^{2}/2(2m)$ exactly. \ Shaded areas as in Fig. 1. 
 } 
\label{fig:tau2} 
\end{figure}

To model cuprate superconductors consider the bosons confined to an infinite
set of planes stacked along the $z$-direction, parallel to each other with
equal spacing $c$ between adjacent planes. \ The BEC transition temperature
formula, for $s=2$ bosons in each plane, is the {\it implicit} $T_{c}$%
-formula \cite{wk} 
\begin{equation}
k_{B}T_{c}={\frac{2\pi n_{B-3D}\hbar ^{2}c}{m_{B}\hbox{ln}%
[M_{B}c^{2}k_{B}T_{c}\nu (t_{c})/\hbar ^{2}]},}  \label{WKeqn}
\end{equation}
and is valid in $2+\epsilon $ dimensions where $\epsilon $ is small and
given by 
\begin{equation}
\epsilon \simeq 2[\hbox{ln}(n_{B-3D}M_{B}c^{3}/m_{B})]^{-1}{.}  \label{eps}
\end{equation}
This expressly vanishes as $M_{B}c^{3}\rightarrow \infty $, as it should. 
Here $n_{B-3D}\equiv N/L^{3}$ while $\nu (t)=1-t+O(t^{2})\simeq 1$ if $t<<1$%
, where $t\equiv \hbar ^{2}/M_{B}c^{2}k_{B}T=\hbar
^{2}/2mc^{2}(M_{B}/m_{B})k_{B}T$. Using $\hbar ^{2}/mk_{B}=88,419\ $ K \AA $%
^{2}$ with $m$ the electron mass, and $c=12$ \AA , this inequality is well
satisfied for the higher cuprate transition temperatures, today ranging up
to 164 K, since $T_{c}>>307$K$/(M_{B}/m_{B})$ as typically $M_{B}/m_{B}$ can
range from $10^{2}$ to $10^{5}$.\ \ Clearly, for\ $M_{B}c\rightarrow 0$
(infinitely separated planes and/or perfect confinement to the $z$-direction
in each plane) $T_{c}$ vanishes as it should in 2D. \ As state, this $T_{c}$%
-equation is {\it implicit} or transcendental, unlike the simpler {\it %
explicit} $T_{c}$ equations (\ref{BEC}) and (\ref{gentc}), and has been used
for varied purposes by numerous authors \cite{Ren,micnas,ranninger}---though
only for {\it quadratic }dispersion bosons.

We have generalized (\ref{WKeqn}) for any $s>0$\ and found 
\begin{equation}
k_{B}T_{c}=C_{s} \left[ {2\pi n_{B-3D}sc}/\Gamma (2/s) g_{2/s}({e^{-\hbar
^{2}/M_{B}c^{2}k_{B}T_{c}}}) \right]^{s/2}.  \label{nB}
\end{equation}
Thus, an exact (again, implicit) equation for $T_{c}$ is obtained for any $%
s>0$. For $s=2$ and $C_{2}=\hbar ^{2}/2m_{B}$ we recover (\ref{WKeqn}). \ In
Fig. 1 (right) we plot results for both values of $s$ as points, for $%
M_{B}/m_{B}=10^{2}$ and $10^{5}$. \ They can be seen to differ very slightly
from the results for $s=2$ or $s=1$ bosons in $d=2+\epsilon $ dimensions,
with $\epsilon $ small, that came directly from (\ref{gentc}) which,
moreover, is valid for all $d>0$.

To compare our results, consider the Berezinskii-Kosterlitz-Thouless \cite
{BKT} transition temperature formula 
\begin{equation}
k_{B}T_{c}^{BKT}=\frac{\pi }{2}\frac{\hbar ^{2}n_{B}}{m_{B}}  \label{BKT}
\end{equation}
valid in 2D, and assume as before that $n_{B}=n/2$ $\equiv k_{F}^{2}/4\pi $
and $m_{B}=2m$. \ This gives $T_{c}^{BKT}/T_{F}=1/8=0.125$ and is displayed
as a square in Fig. 1. \newline

In conclusion, BEC $T_{c}$'s related to a pure gas of unbreakable composite
linear-dispersion bosons were calculated in $d=2$ for Cooper pairs formed
via a general separable potential and whose coupling for any CMM is
characterized solely by its two-body binding energy in vacuum. \ A $T_{c}$%
-formula valid for {\it any }dispersion relation of the form $\varepsilon
_{K}=C_{s}\,K^{s}$ with $s>0$ and $C_{s}$ constant is deduced for the BEC of
a pure gas of unbreakable bosons confined to move in infinitely-many
identical planes parallel to each other. \ It is a peculiar {\it implicit }$%
T_{c}$-equation valid in $2+\epsilon $ where $\epsilon $ is small and
vanishes, as it must, when the inter-plane spacing or the boson mass in the
direction perpendicular to the planes diverges. \ However, for either $s=2$
or $1$ we find results that are just slightly different than those of the {\it %
explicit} BEC $T_{c}$-formula which is valid more generally for any $d>0$. 

Partial support from UNAM-DGAPA-PAPIIT (M\'{e}xico) \# IN102198, CONACyT
(M\'{e}xico) \# 27828 E, DGES (Spain) \# PB95-0492 and FAPESP\ (Brazil) is
acknowledged.

\end{document}